# Nanoindentation in multi-modal map combinations: A Correlative Approach to Local Mechanical Property Assessment

Christopher M Magazzeni*[1], Hazel M Gardner[1], Inigo Howe[1], Phillip Gopon[2], John C Waite[1], David Rugg[3], David E.J. Armstrong[1], Angus J Wilkinson[1]

[1] Department of Materials, University of Oxford, UK - Parks Road Oxford OX1 3PH

[2] Department of Earth Sciences, University of Oxford, UK - S Parks Rd, Oxford OX1 3AN

[3] Rolls-Royce plc, Derby, UK - Elton Rd, Allenton, Derby DE24 8ED

*Corresponding Author: christopher.magazzeni@materials.ox.ac.uk

**Graphical Abstract**

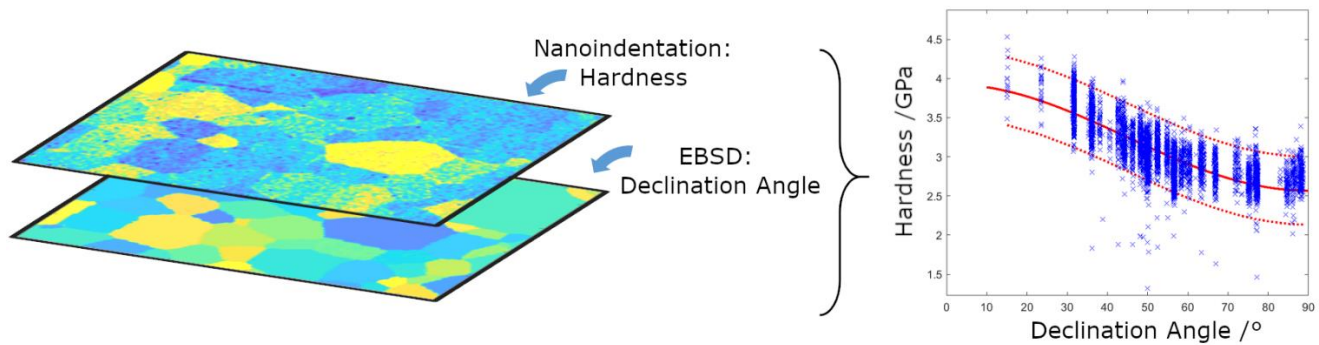

## Abstract

A method is presented for the registration and correlation of property maps of materials, including data from nanoindentation hardness, Electron Back-Scattered Diffraction (EBSD), Electron Micro-Probe Analysis (EPMA). This highly spatially resolved method allows for the study of micron-scale microstructural features, and has the capability to rapidly extract correlations between multiple features of interest from datasets containing thousands of datapoints. Two case studies are presented in commercially pure (CP) titanium: in the first instance, the effect of crystal anisotropy on measured hardness and, in the second instance, the effect of an oxygen diffusion layer on hardness. The independently collected property maps are registered using affine geometric transformations and are interpolated to allow for direct correlation. The results show strong agreement with trends observed in the literature, as well as providing a large dataset to facilitate future statistical analysis of microstructure-dependent mechanisms.

## Introduction

Over the past decades, significant advances have been made in techniques used to determine material properties on an ever finer length scale, allowing for a better understanding of basic material property-microstructure relationships [1]–[3]. These developments have provided crucial insight into fundamental materials science, facilitating more accurate multi-scale modelling [4]–[10], as well as the necessary property assessment for modern manufacturing processes with small features [11]–[16].

Researchers have used crystallographic techniques [3], [17]–[20], composition measurements [21]–[25], and nanoindentation [21], [26], [27] amongst a variety of other techniques to determine local properties. As microstructural feature sizes become smaller, and systems more complex, there is a need for a distributed approach to understanding how multiple factors may affect one another, and in turn affect the viability of parts in complex service environments [21], [28].

Recent developments in nanoindentation instrumentation have enabled a wide range of experimental procedures, including more complex loading and sensing regimes such as strain-rate experiments to measure the effect of indentation rate [29]–[31], and more complex specimen preparation techniques such as micro-pillar compression to isolate microstructural features [32]–[36]. A recent addition to the tools available for local mechanical property assessment has been the development of rapid nanoindentation map-ping: mechanical property maps of hardness and modulus can be obtained that resolve micron-scale microstructural features [37]–[39], and create, at an unprecedented rate, large datasets for statistical analysis [40].

Work carried out using this technique has shown developments in distinguishing between phases in cement [40], [41], or dissimilar material coatings [38] through indentation alone and has enabled the collection of statistically significant datasets [38], [41]–[46]. Statistical clustering methods are available to classify data points and extract characteristic properties from a finite set of discrete phases. However, in microstructures where changes in local property are continuous, or are within a few tens of percent, alternative approaches requiring additional source signals need to be developed in order to de-convolute more nuanced structure-property relationships. There is scope to collect multi-dimensional datasets from microstructurally-rich materials that combine chemical, crystallographic, and mechanical data at high resolution. However, the challenge remains to correctly collect, correct, align, and correlate these very different properties.

Nanoindentation mapping contains an inherent trade-off between resolution and accuracy, manifested as the trade-off between indent spacing and depth. Deeper indents generally provide higher accuracy data of material bulk properties [47] though care must be taken when spacing indents close to one another as the plastic zone of one indent can affect the result of the second. This depth to spacing ratio has been discussed by Sudharshan Phani *et al.* [37], where a ratio of depth to spacing of 1:10 is recommended in nanoindentation maps. This directive on depth to spacing ratios alone does not imply any resolution limit in nanoindentation mapping. However, in combination with the above statement on shallow indent accuracy, the trade-off is: the smaller the indentation spacing, the larger the error in the data. This is either due to depth to spacing ratios approaching or lying below the above recommendation, or due to surface effects at shallow depths affecting bulk material hardness measurement [47]. Some experiments indicate that for nanoindents approximately 100 nm in depth, this error for each indent is larger than the error from decreasing the depth to spacing ratio, justifying the use of a 1:7 ratio for certain experiments where high spatial resolution is desired.

In this paper, we present a method for directly correlating crystallographic data from Electron BackScatter Diffraction (EBSD), chemical data from Wavelength Dispersive Spectroscopy (WDS) of X-rays, and mechanical properties data from nanoindentation. From this we study variations in mechanical response as a function of crystal orientation, interstitial content (in the case presented, oxygen content), and spatial position, rather than simple characteristic properties for a small number of discrete phases. We illustrate the method and analysis possible using titanium test specimens, showing the first steps in rapidly obtaining structure property relations from complex systems at high resolution. We present two commercially pure titanium systems to be mapped and correlated: the first with a bulk un-textured alpha microstructure, and the second with an oxygen surface diffusion layer.

### Results

#### Dataset 0 – Nanoindentation comparisons and tip calibration persistence

Figure 1 shows data from the nanoindentation map performed on the fused silica reference material, illustrating tip calibration persistence. The map, where each indent corresponds to one pixel, was constructed of 3x3 bundles, starting from the bottom-right of the image, and progressively moving up in a serpentine motion starting in the negative x-direction. A graphical representation of this strategy is provided in the supplemental information in Figure 9.

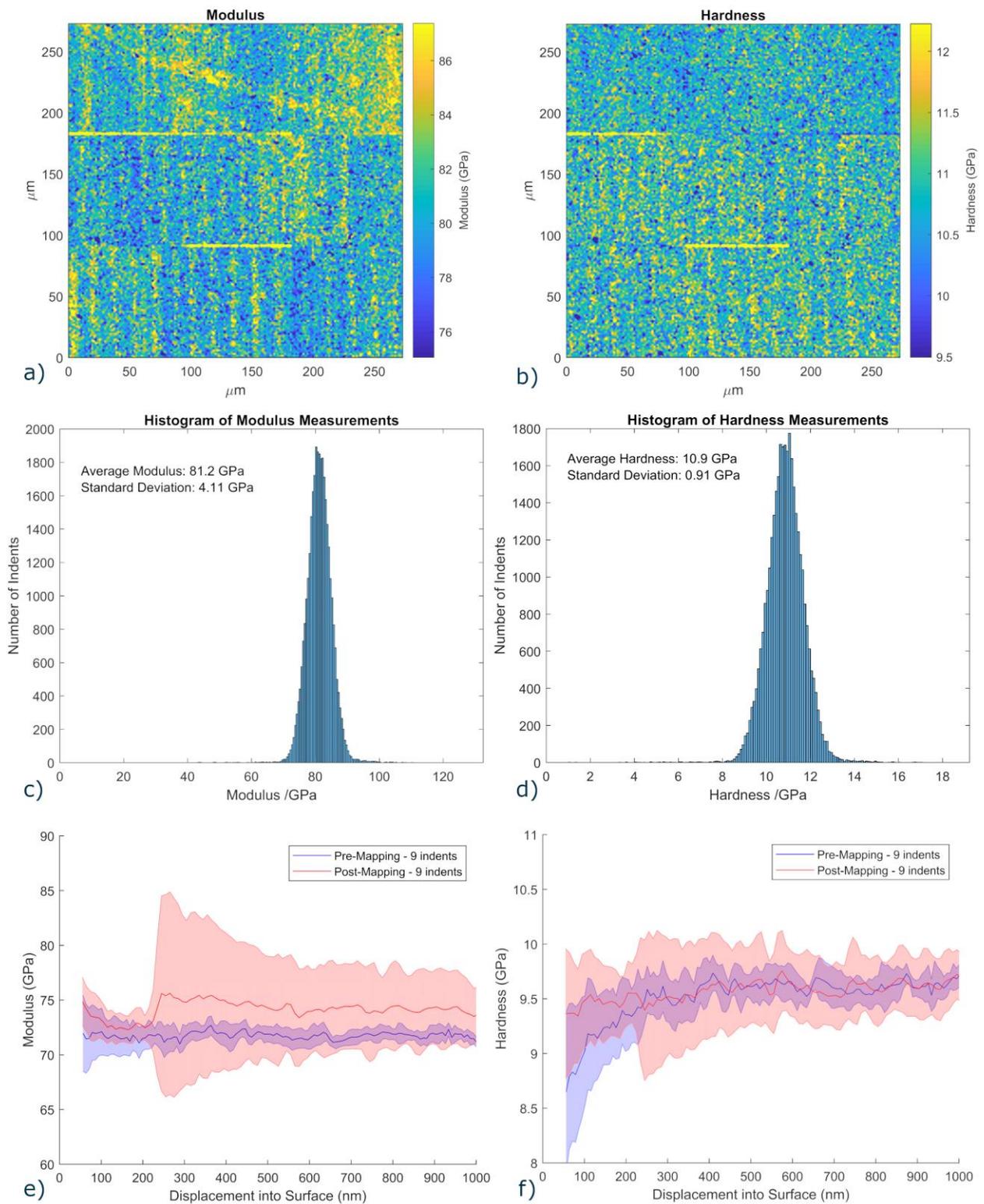

*Figure 1: Hardness (a) and modulus (b) figure of nanoindentation map on Fused Silica. Indents are performed at 3mN, spaced 1.5um apart. This corresponds to a depth/spacing ratio of 1:10, with indents approximately 150nm deep. Histograms of modulus (c) and hardness (d) maps obtained in (a) and (b). CSM nanoindentation data corresponding to modulus (e) and hardness (f) both before (blue) and after (red)*

In Figure 1 a) & b) it can be seen that there is an inaccuracy in the stage movement travelling between bundles: the horizontal lines of high modulus correspond to indents more closely spaced than programmed. The collection strategy within bundles can also be seen, with vertical stripes corresponding to systematic errors in x-stage positioning within bundles. These factors contribute to noise in the nanoindentation mapping, but do not prevent the extraction of trends from the full map.

Histograms of the modulus and hardness values for the entire map are shown in Figure 1 c) & d). The mean modulus is 12.5% higher than the value of 72 GPa established using slower standard CSM mode nanoindentation (more discussion of this offset in hardness will be given later). However, it is reassuring that there is minimal increase in modulus or hardness across the bundles, indicating little tip wear within a map of this size.

The CSM data shown in Figure 1 e) & f) supports this conclusion. The first set of nine indents made before mapping were used for calibration, while the second set obtained after mapping show little to no change in modulus, ~2 GPa, at the 150 nm depth used for the indentation mapping. This indicates there is no significant systematic change in tip calibration during the indentation mapping.

Further to discussions of tip calibration persistence, it is necessary to consider the spatial resolution of this mapping technique. We compare results obtained using conventional CSM nanoindentation measurements and nanoindentation mapping on the same material. Figure 10 and Figure 11 in the supplementary information provides an example of the discrepancy between nanoindentation mapping and CSM nanoindentation results for a fused silica calibration specimen, and the CP titanium specimen studied. There is a persistent artificial increase in hardness and modulus recorded when performing nanoindentation mapping, which decreases as depth is increased. This may be related to the rate at which the indents were performed, or due to the difference in the way these values are calculated. Despite these systematic variations, nanoindentation mapping provides data in line with literature accepted values for fused silica.

**Dataset 1**

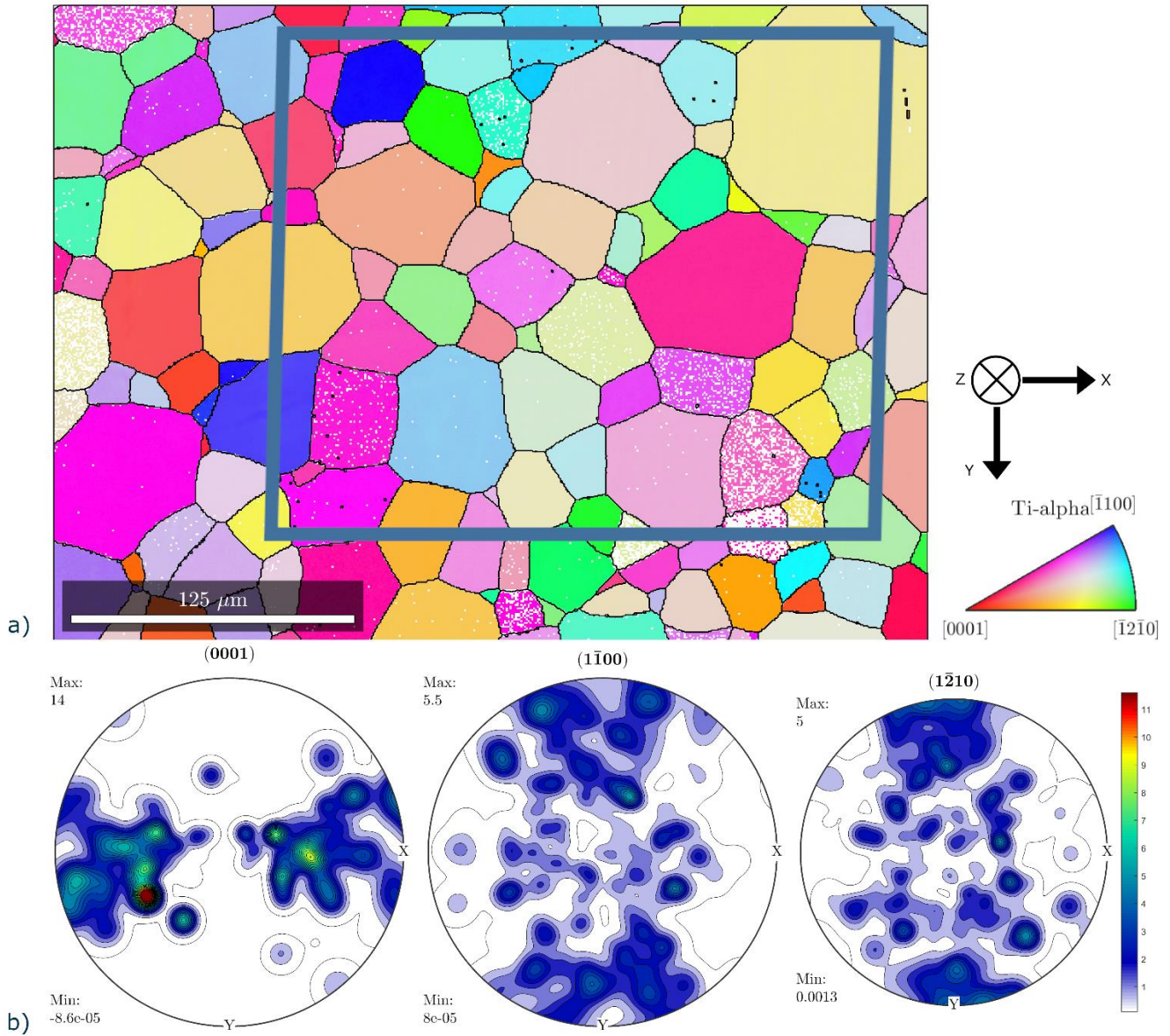

*Figure 2: a) IPFZ EBSD map of the CP titanium specimen with pole key and b) pole figure. Overlay indicates the approximate area to be nanoindented.*

Figure 2 shows EBSD data obtained from the commercially pure titanium specimen which has grains with a mean size of approximately 15 μm with weak texture allowing for a full range of orientations to be probed. A nanoindentation map was then performed on the specimen within the region mapped by EBSD, with results shown in Figure 3.

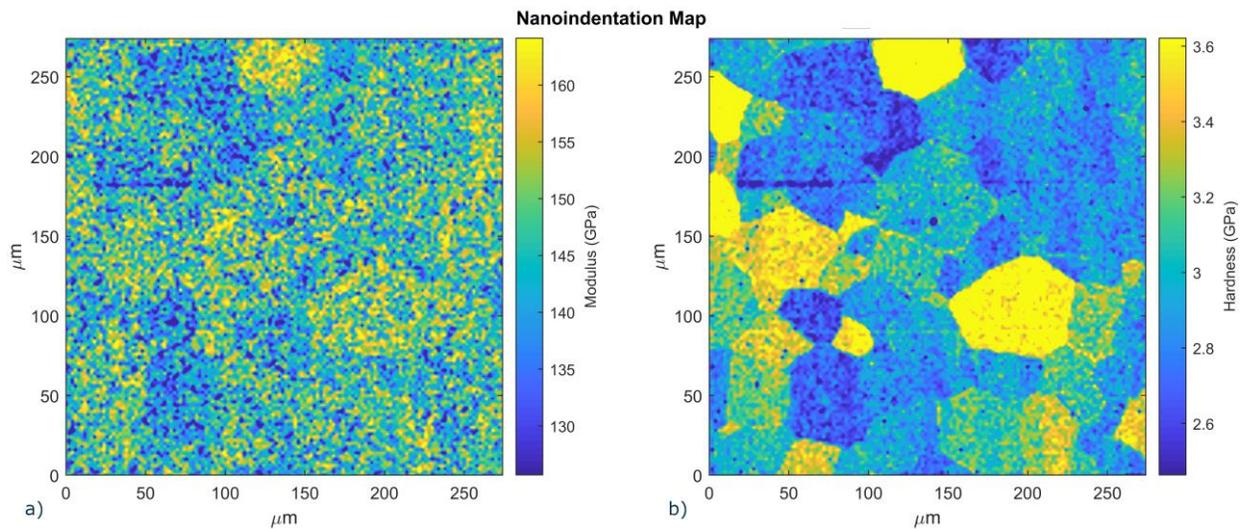

*Figure 3: Nanoindentation map of CP titanium showing a) modulus and b) hardness. The indentation map was performed at a fixed load of 3mN corresponding to an approximate depth of 200nm, and an indent spacing of 2um. The map is 138x138 pixels or indents in size, corresponding to 19044 data points.*

The hardness map in Figure 3 reveals significant contrast between grains, whereas there is lower grain-to-grain contrast in the modulus map, relative to the noise. We note the dissimilarity in contrast between hardness and modulus maps in Figure 3, in contradiction to previous reports of this well-studied material [20], which points to a more complex relationship between the two. As a consequence, the rest of this paper discusses only the hardness data in order to remain confident in the trends observed. This difference between the hardness and modulus maps is discussed further in the nanoindentation mapping methods section.

An affine transformation was performed in order to correctly align the EBSD map to the nanoindentation map using a set of eight triple junction points in these maps, as highlighted in Figure 12 (supplementary information).

Once aligned, the two maps can be interpolated to the same size, so that correlations between various property fields can be undertaken easily. To down-scale the EBSD map, nearest-neighbour interpolation is used, a decision which is discussed in Limitations and Methods. For consistency, we use the same colour scheme for EBSD derived maps as for nanoindentation derived ones.

The crystal orientation data is readily simplified to display maps of the declination angle between the basal plane normal (i.e. the c-axis) and the surface normal (i.e. indentation direction), as seen in Figure 12 (supplementary information). This is done as commercially pure titanium crystals are well understood to have high crystal anisotropy due to changes in the declination angle [20], [48], [49]. In Figure 4 a) and b), the map of declination angle has been aligned to the nanoindentation map, allowing direct comparison between the two maps since the spatial sampling is now equal.

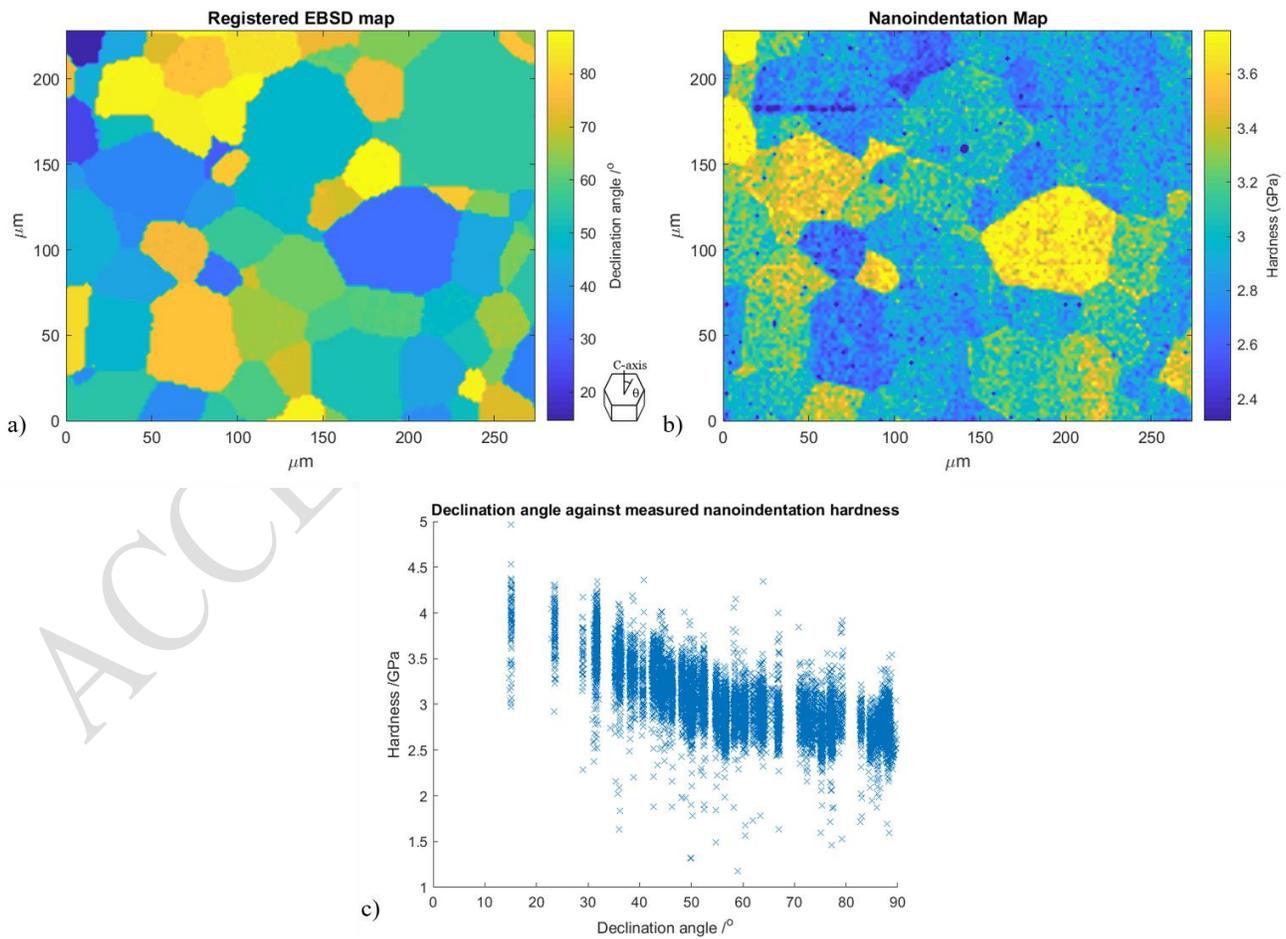

These arrays can now be used to create scatter plots relating pixel-by-pixel the declination angle of the c-axis to the measured hardness, shown in Figure 4 c).

This curve closely resembles that reported in previous experiments, notably by Britton *et al* [20]: the anisotropic effect of declination angle on measured nanoindentation hardness is clearly evident, and the range of hardness (~1GPa) can be used to validate these results. The benefit of this technique lies beyond the immediately visible: a very large quantity of data (>10,000 data points) has been collected in a relatively short time, while retaining all spatial and property information.

The data can be seen to comprise primarily of vertically spread populations at fixed declination angles. Each column of data points in the plot in Figure 4 c) relates to points with the same orientation, often referring to a singular grain but occasionally multiple grains with the same shared orientation. Within these grains, there is then a spread of hardness data, arising from other contributing factors to hardness, as well as measurement noise.

There are two primary contributions to the noise in this dataset: the misattribution of pixels between maps, and the physical influence of grain boundaries on nanoindentation hardness. The first is a result of the error in transforming the EBSD data and aligning it onto the hardness map, which will be discussed further in Discussion and Limitations. This error will occur in points lying close to grain boundaries: when overlaid, some pixels may contain the hardness data of one grain but the EBSD data of the adjacent grain due to misalignment. The second is due to the real effect that indents close to grain boundaries respond differently to indents performed in the centre of grains. Plastic zones that impinge upon grain boundaries can cause dislocation pile-up and slip transmission, or the grain boundary might act as a dislocation source/sink itself, strongly affecting measured hardness [50]–[56]. In order to avoid the influence of grain boundary associated variation, and capitalising on the wealth of data available, points can be isolated based on spatial location relative to the microstructure. The EBSD data contains information on grain boundary location, and this can be used to calculate the distance of every indent to its nearest grain boundary, shown in Figure 5.

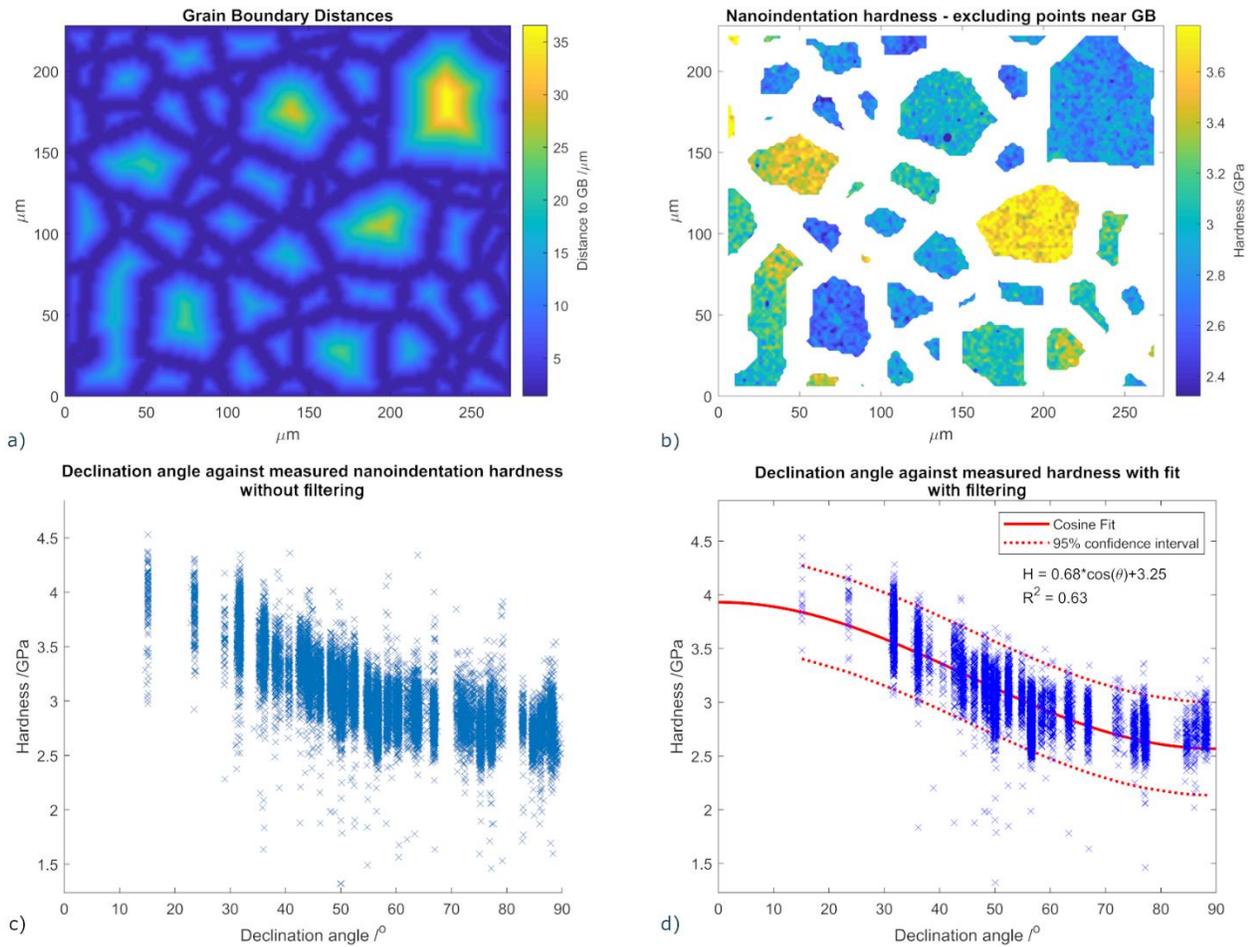

*Figure 5: a) Distances to grain boundaries obtained through the EBSD map, b) the nanoindentation obtained hardness map excluding points in (a) with a value below 7 µm, c) the scatter plot of declination angle, measured through EBSD, plotted against hardness measured through nanoindentation mapping as seen in Figure 4, and d) the scatter plot of c), excluding all points that lie within 7um of a grain boundary as determined by EBSD with an arbitrary simple function fitted. 6246 data points are shown in c). A 95% confidence interval is shown in the plot.*

The above scatter relationship can then be plotted excluding all points that lie near grain boundaries – this ensures that points are only plotted if they can confidently be assigned their orientation, and are unlikely to be impacted by the nearby grain boundary.

Figure 5 d) shows the scatter plot excluding all points within 7 μm of a grain boundary, with Figure 5 b) showing the remaining points on the original nanoindentation map. These figures contain 6246 data points, representing still a significant dataset, and the trend in Figure 5 d) is significantly clearer than in c).

A simple, arbitrary functional form that fits the boundary conditions of the data has been selected and used to fit the data in Figure 5 d) and is found to give a reasonable representation of the trend. It can be reasonably said that the remaining scatter within the data is largely a consequence of measurement noise in nanoindentation data at about 10%. This fitted relationship can be used improve the spatial alignment of EBSD and nanoindentation datasets. The measured nanoindentation hardness values were used with the fitted relationship to produce a 'simulated' declination angle map. The simulated map does not contain all the crystallographic information within the EBSD data, but abrupt changes in the effective declination angle clearly demark some of the grain boundaries in a similar way to that evident in the EBSD data. A Sobel edge detection filter [57] was applied to the EBSD, and hardness-derived simulated declination angle maps, to produce the maps shown in Figure 13 (supplementary information).

These were used within an enhanced correlation coefficient (ECC) cross-correlative image alignment procedure [58], [59] to determine a homographic correction, allowing for two more degrees of freedom and capturing a greater range of distortions [60]. This further correction to the transformation from the spatial frames of the EBSD to nanoindentation data is now based on the extent of over-lapped fields, rather than a small number of discrete user-defined control points. The scatter plot and best fit relationship through the hardness and EBSD measured declination angle shows only very minor changes for points remote from the grain boundaries after this correction to the spatial alignment. From this, the same structure-property scatter plot can be obtained with more confidence in the assigned orientation for each point.

Following this correction to the spatial alignment, the behaviour near grain boundaries can be explored with more confidence. The EBSD-measured declination angle at each pixel is used within the fitted relationship to determine the hardness expected for a grain in this orientation. The measured hardness map and the expected hardness calculated from crystal orientation are compared in Figure 6 a) & b). This expected hardness is based on a fit to data obtained well away from grain boundaries, and benefits from lower noise due to the smoothing/averaging inherent in the data fitting process.

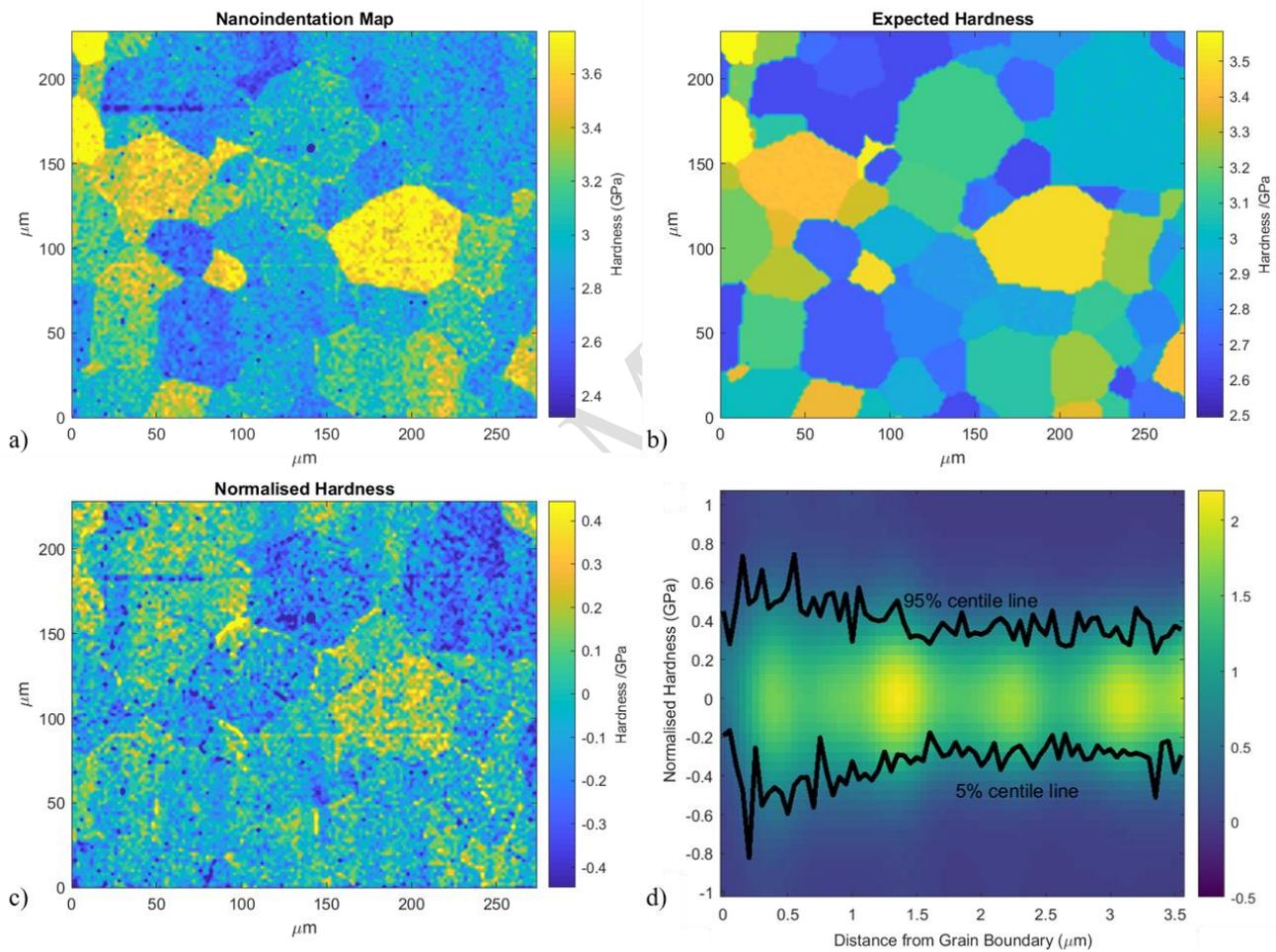

*Figure 6: a) nanoindentation measured hardness, b) expected hardness through the registered EBSD dataset and obtained relationship, c) normalised hardness through subtraction of a) and b), d) normalised hardness as a function of distance to the grain boundary. The colour scale in d) indicates log(1+N) where N is the number of indents at each combination of normalised hardness and distance from a grain boundary, while the lines show the 5th and 95th percentile values from normalised hardness distributions for indents at different distances from grain boundaries.*

Variations in measured hardness across the microstructure can be made more apparent by subtracting the expected hardness from the measured hardness to give what will be referred to as the normalised hardness. This removes the quite marked (>1 GPa) grain-to-grain hardness variation caused by orientation (Figure 6 a) and b)), leaving more subtle small variations visible in the normalised hardness map (Figure 6 c)). In this case, we expect only secondary signals: noise, grain boundary effects, and possible chemical hetero-geneities (though these have not been shown to be of concern in CP titanium).

Grain boundaries are clearly highlighted in the normalised hardness map (Figure 6 c)). Some segments show higher measured nanoindentation hardness, while others show lower measured hardness. There are clear effects from noise, as well as from small deviations in nanoindentation positioning: horizontal stripes can be seen, resulting from overlapping indents and their resulting plastic zones due to stage misalign-ment. Some individual grains have a slightly higher mean normalised hardness, and others lower. Possible explanations for this variation include the approximate form of the assumed fitting function, as well as explanations with a physical basis, such as interactions with close sub-surface grain boundaries, or grains that may have high residual stresses due to cooling [28]. It is clear however, that a primary source of variation in the normalised hardness is due to grain boundary effects. This can be visualised, as shown in Figure 6 d), as normalised hardness as a function of the distance to the grain boundary, showing an in-creased divergence in normalised hardness in the vicinity of the grain boundary. The details of this, and further results, will be discussed in a further publication.

### Dataset 2

To demonstrate further the capabilities of this method the second example also includes chemical infor-mation, and is a study of an oxygen diffusion layer generated on a CP titanium specimen that had been subjected to 230 hours at 700°C in air (specimen 2). Figure 7 shows, from left to right, the EBSD obtained declination angle map, the EPMA obtained oxygen concentration map, and the nanoindentation hardness map. All maps have been registered using four user supplied control points to define two affine transformations mapping the EBSD and EPMA spatial data to that of the nanoindentation map.

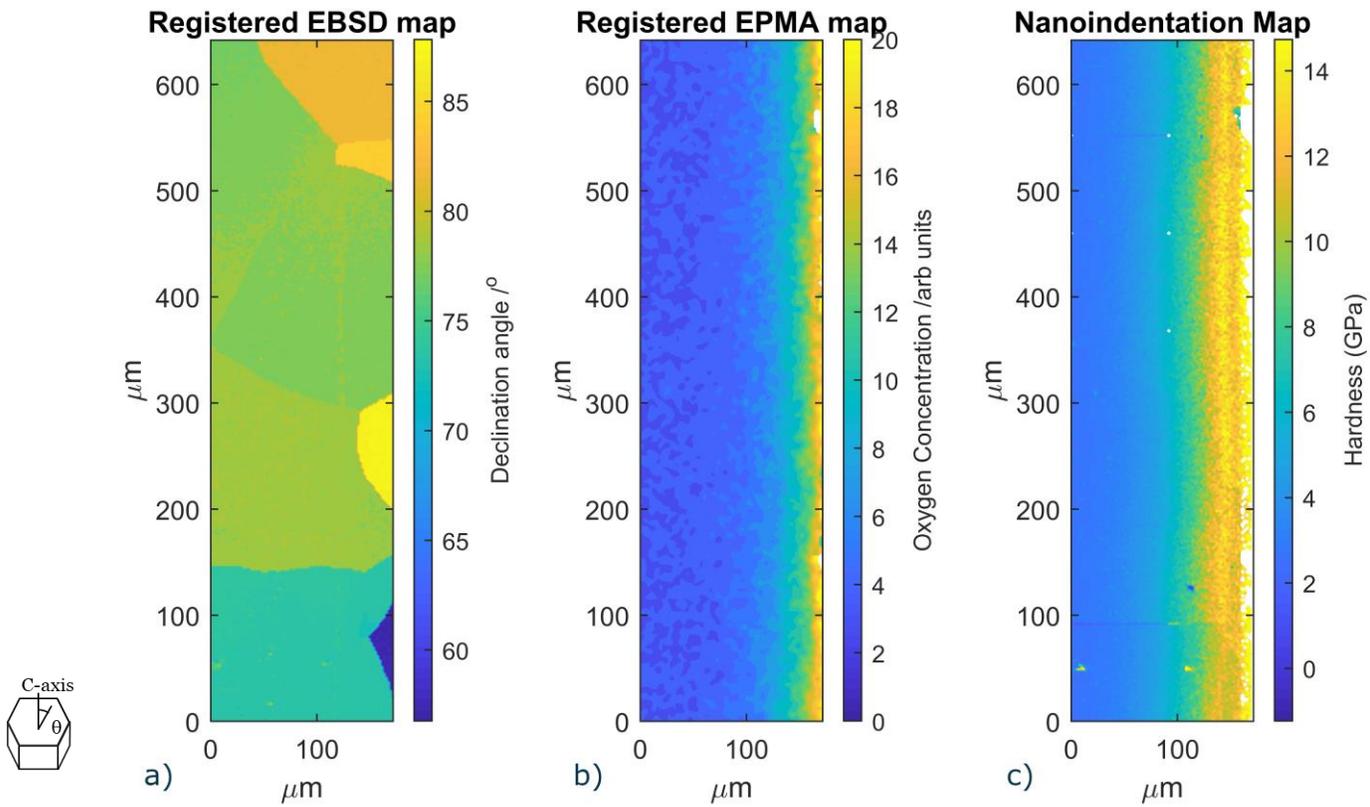

*Figure 7: Dataset 2 represented as: a) EBSD map showing declination angle, b) EPMA map showing oxygen concentration, and c) nanoindentation map showing measured hardness.*

The EBSD map shows that there is a strong texture element in this region of the specimen: there is a narrow distribution of grain declination angle. This keeps the specimen largely at the same orientation, allowing collection of a large dataset to study the relationship between hardness and oxygen content. Similarly to specimen 1, it remains possible to observe the relationship between the nanoindentation hardness and the declination angle. Figure 14 (supplementary information) compares data found in this near surface layer to the trendlines established for specimen 1.

These data show that a low number of orientations are probed within this region, and that there is a significant spread of hardness at a given orientation, with hardness reaching very high values of 20 GPa – four times the lower range of data points which are more in-line with hardness levels measured in sample 1. The reason for this spread towards higher hardness values, clearly, is that this graph omits any information regarding oxygen content. The EPMA map in Figure 7 shows that data points with high oxygen content can be excluded by considering points deeper than ~120 μm. For this subset the average hardness is 3.0 GPa much more in line with the data presented in Figure 5 d) for the bulk specimen 1, though with declination angles restricted to the range 60°-90°.

Since both oxygen content and grain orientation vary, it is helpful to create a 3D plot of hardness versus oxygen content (from EPMA) and declinations angle (from EBSD) as is shown in Figure 8.

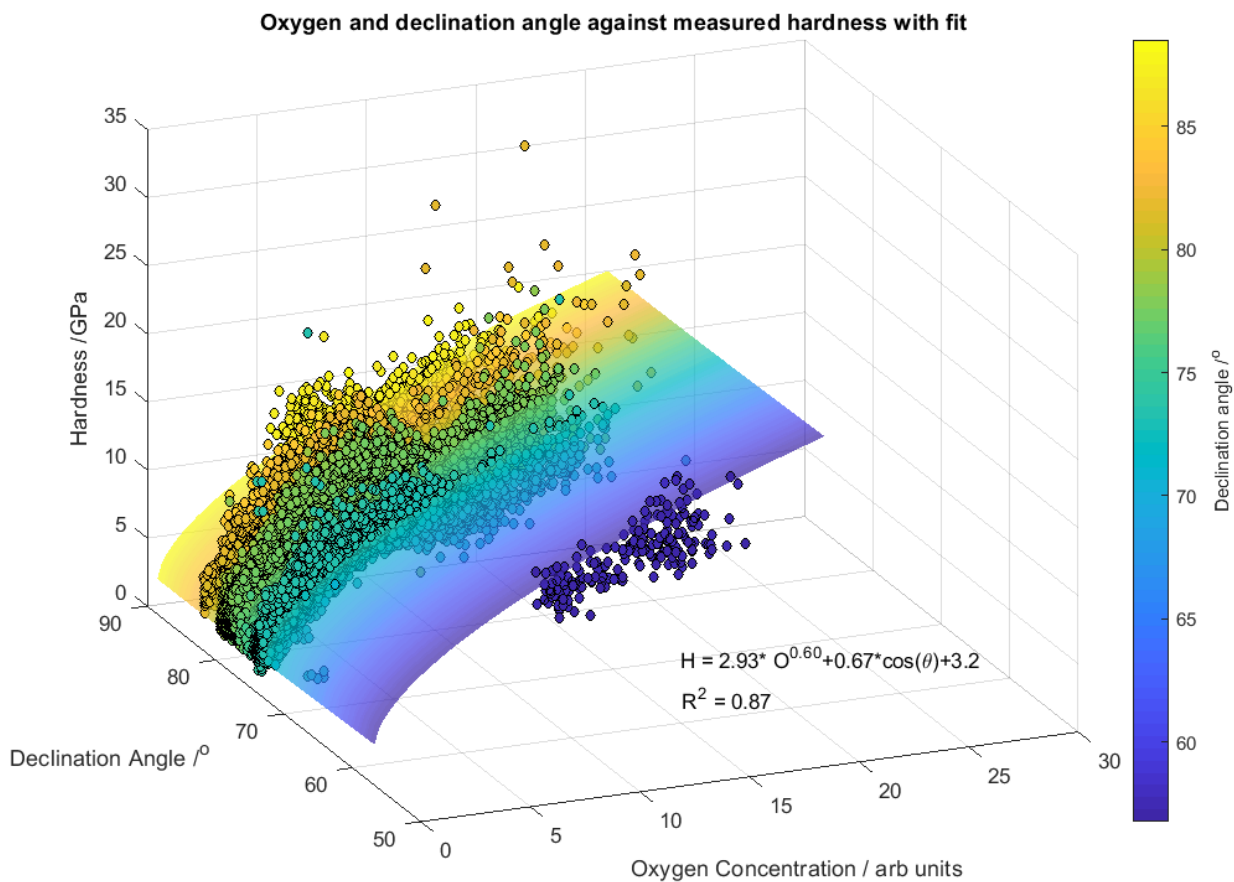

Figure 8: EBSD obtained declination angle against EPMA oxygen abundancy against measured nanoindentation hardness. A total of 22586 data points are shown. A fit with a combination of the orientation component obtained from Dataset 1, as well as a power exponent fitted via least-squares regression, is included. For ease of visualisation, points (and fitting surface) are coloured by the declination angle.

It is clear from Figure 8 that the spread in hardness in Figure 14 is much more significantly correlated with the elevated and varying oxygen concentration, rather than crystal declination which is restricted to a relatively limited range, and is known from Figure 5 d) to have a smaller effect on hardness. This graph shows that hardness increases as a function of oxygen concentration, as described in the literature [61], and can be fitted with a power law function as done in Chen *et al* [62]. A unique feature of this curve, is the presence of a dip in hardness at approximately 13 units of oxygen measured. As we retain information on spatial information of each of these points, we can locate every point that lies below the 95% confidence interval, shown in Figure 16 (supplementary information). These points all lie in a narrow band near the surface, where the oxygen concentration is highest. The reason for this dip in hardness is unclear, but it could be a surface effect, or could be linked to the higher oxygen concentration in this region. The scatter plot of oxygen concentration against hardness is shown in Figure 15 (supplementary information).

The ability to preserve, for every point, both spatial information and orientation information allows us to separate these source signals. In the case of grains, we can observe oxygen – hardness relationships for each set of orientations individually, shown as the different colourations in Figure 8. We can also understand the clustering of data points in this plot as the inherent spatial positioning of the grains: notably the grain at 58° of declination angle does not extend beyond approximately 30 μm from the surface, leading to no points of that orientation having bulk-levels of oxygen.

It can be seen that this technique rapidly provides us with orientation informed oxygen diffusion layer hardening information for this CP titanium system via the collection and alignment of three maps. Use of this technique can be extended to study more complex microstructural and chemical systems, enabling industrially motivated questions to be answered. For example, work by Gardner *et al* [63] uses correlative hardness mapping, EPMA, EBSD and atom probe tomography to quantify the effect of oxygen ingress on micromechanical properties of an in-service jet engine compressor disc.

## Discussion and Limitations
### Reliability of input information

The rapid indent testing (3.5 sec/indent) used for the nanoindentation mapping generates some limitations to the data generated. The most significant of these is the current limitation to analysis of hardness data. This decision was taken in part due to some further testing required to understand the details of the modulus calculations, as well as the lack of modulus contrast in a material that is known to produce these variations. Hardness data from nanoindentation mapping also contains a significant noise level with variations exceeding 10% of the average value obtained, likely due to the rapid nature in which it is collected.

The spatial alignment of nanoindentation mapping must also be considered. It can be seen from the resulting maps that rows of indents near the edge of a bundle occasionally overlap with neighbouring bundles, as a result of small stage misalignment. This is particularly noticeable in high spatial-resolution nanoindentation mapping, and can adversely affect data quality. Along these rows of indents, there can be significant changes in depth-spacing ratios, and in extreme cases include indents overlapping entirely, with an example shown in Figure 17 (supplementary information). These effects are typically seen when the indent spacing approaches the accuracy limit of the stepper motor used to drive the larger movements of the sample between bundles. Where the stepper motor makes slightly larger spacing between bundles, this does not influence the hardness values but will cause a slight dis-registry with spatial locations in the other maps.

A discussion of the details and limitations of the other input methods used in this paper can be found in previous publications related to those methods in particular [19], [64]–[68].

### Reliability of analysis

The limitations of the analysis can also be split into two categories: alignment and interpolation. The first is an issue of error arising from the user input of reference points. A small change of a few pixels in one of the reference points can lead to significant changes to the affine transformation, particularly at the corners of the map and when maps have a high aspect ratio. This could lead to the misattribution of data points as being in the wrong grain. In an attempt to quantify the extent of such issues, the EBSD and nanoindentation maps of dataset 1 were registered multiple times using freshly input control points marked by the same observer. On average $4.3 \pm 1.0$ % of pixels were assigned different grains when pairs of registered datasets where compared. When the additional full field ECC alignment was used then $3.9 \pm 0.5$ % of pixels were assigned to different grains.

The second is the question of interpolation. Datasets with discrete boundaries, such as in EBSD, must be interpolated with "nearest-neighbour" methods in order to avoid the fabrication of non-existent orientation data: grain boundaries are discrete, and a point interpolated as any average of points across the boundary is unphysical. However, this discrete nature of the grain assignment often causes stepped or serrated grain boundaries, and amplifies the errors in alignment discussed above.

Despite these limitations, the method presented provides a large quantity of data which can confidently be said to be well assigned, showing trends with high fidelity and data quantity.

### Quantitative results

The structure-property relationships obtained show strong agreement with those collected and discussed in the literature before. Britton *et al.* [20] have reported the extent of anisotropic response of nanoindentation in CP titanium, with largely similar results in hardness data. However, as mentioned, the values of hardness and modulus collected via nanoindentation mapping have yet to be robustly compared to conventional CSM data, which may justify the relative difference in hardness values as well as our lack of modulus data.

The relationships obtained between nanoindentation hardness and oxygen, fitted with the same power law function as described in Zheng *et al* [62], broadly agree with the literature [61], [69]. Without exact quantification of the oxygen signal it is impossible to draw further conclusions from this, and further work aims to obtain reference points for these curves in order to quantify absolute measurement of oxygen weight percentages.

Finally, secondary effects such as grain boundary hardening can be seen, much in agreement with experiments in the literature showing variation in nanoindentation hardness in the vicinity of grain boundaries up to several microns in distance [52], [54], as well as the sensitivity of this variation to the crystal orientation of both grains [55], [56], [70], [71]. Results in the further analysis of these results to determine the underlying mechanism for this will be discussed in further publications.

### Conclusion

Nanoindentation mapping has recently become an available tool for the collection of a large number of indents in a short time (e.g. dataset 1 - 15870 points in 15 hours). This has allowed for spatially resolved maps displaying mechanical properties, and providing datasets that are larger than have been previously obtainable.

Current literature showcases the utility of nanoindentation mapping through examples illustrating the ability to separate discrete phases and obtain statistically significant datasets of hardness and modulus for individual phases with large difference. This is work which otherwise would have required a targeted approach. However, until now there has been a challenge in determining results from continuous variables, such as crystallographic anisotropy, as it requires a separate source signal to be collected and properly registered.

This work demonstrates that when nanoindentation mapping is used in combination with other source signals, and correctly registered, rapid collection of structure property relationships with more subtle or continuous variables can be obtained.

Furthermore, we have shown that it is possible to deconvolve the effect of multiple varying structural properties on the mechanical response to nanoindentation, and clearly observe secondary effects such as grain boundary hardening. This paper presents the method for doing so, as well as highlighting the current limitations of nanoindentation mapping, the work required for registration of separate source signals, and the results obtainable with two standard use cases. Future work will endeavour to explore the more subtle variations observed, such as grain boundary hardening, as well as examine more complex use-cases with strong industrial interest, such as oxygen diffusion in titanium alloys.

## Methods

### Sample Preparation

Some preliminary measurements (dataset 0) were obtained from a fused silica sample, often used as a calibration material system for nanoindentation. The sample had been supplied by Goodfellow Ltd. and was used as polished.

Two specimens of grade 1 commercially pure titanium, supplied by Timet UK Ltd. were prepared for this experiment with an approximate size of 5x5x10 mm. The chemical composition of these, in weight per-cent, were: 0.07 oxygen, 0.035 iron, 0.012 carbon, 0.0035 nitrogen, and the remainder titanium [72]. For Specimen 1, experiments were performed without further heat treatment to the material. For Specimen 2, one face was polished to 4000 grit with silicon carbide paper in order to create a uniform surface, and the specimen was subjected to 700°C for 230 hours in lab atmosphere to form an oxygen diffusion layer [73], [62]. Following this, Specimen 2 was cross-sectioned with a diamond saw blade in order to expose the oxygen diffusion profile.

Both specimens were mounted in Bakelite [TM], progressively ground to 4000 grit with silicon carbide paper, and finally polished with a colloidal silica suspension until clear grain contrast became visible under a polarized light microscope. In the case of Specimen 2, the edge which was to be examined was polished as the trailing edge, in order to ensure good edge-retention.

In Specimen 1 a region of good quality polish towards the centre of the specimen was analysed, and in Specimen 2 the region analysed was selected at the surface such that it contained the full oxygen diffusion profile. Each region of interest was demarcated by nano-indents or otherwise recognisable features, and maps were taken of the same approximate region in the following order: an EPMA map, an EBSD map, and a nanoindentation map. This sequence was chosen to minimise the effect of one measurement method onto the other: nanoindentation mapping as the most surface-destructive was performed last, and EPMA which can be affected by the deposition of carbon contamination during EBSD was done first. The carbon contamination deposited via electron beam has been shown in preliminary experiments not to significantly affect nanoindentation readings performed at a depth of at least 100 nm.

### EPMA mapping

EPMA maps for Specimen 2 were acquired on a CAMECA SX5-FE, located in the Oxford Dept. of Earth Sciences, equipped with 5 wavelength dispersive X-ray detectors. Conditions used were: 10 keV accelerating potential, 15 nA beam current, 0.1 second dwell time per pixel, and 500 nm step size. Liquid nitrogen was used throughout the analysis to keep the levels of carbon contamination down. This was done not only to decrease the issues with C contamination for the subsequent EBSD and nano-indent maps, but carbon build-up during long analyses has been shown to adversely affect the EPMA signal at the low accelerating potentials used here [65]. To further keep contamination level down while maximizing the X-ray signal, maps were not acquired with a background pass. Background corrected test maps conducted on regions adjacent to the final maps showed no change in the background signal for any of the X-rays of interest.

X-ray signals collected were: O-Kα simultaneously collected on two WDS spectrometers using PC0 diffracting crystals, Ti Kα simultaneously collected on two WDS spectrometers using LPET diffracting crys-tals, and N-Kα collected using the PC2 diffracting crystal. O-Kα and Ti-Kα signals collected simultaneously were summed to increase the signal-to-noise. Note that we only report relative differences in composition due to the oxide surface layer present on the sample surface.

### EBSD mapping

EBSD data was collected on a Zeiss Merlin FEG-SEM at a beam energy of 15 & 20 keV for specimen 1 and 2 respectively, with a probe current of 10 nA. A Bruker e-Flash HR EBSD detector operated by Esprit 2.0 software was used acquire the EBSD maps with patterns collected and saved at a resolution of 150 x 150 pixels. For specimen 1 the mapped area was 460 μm by 620 μm at a step size of 0.8 μm, while for specimen 2 a 420 μm by 1470 μm region was mapped at 1 μm spacing. The resolution and sampling density of both EPMA and EBSD is sufficiently high to identify relevant features such as grain morphology and oxygen profiles, and these are collected such that they are higher resolution than the nanoindentation maps. A discussion of this will follow in the analysis section.

### Nanoindentation mapping

An Agilent Technologies (now KLA Tencor) G200 nano-indenter system was used to collect nanoindentation maps using the Express Test option. This instrument can perform 1 indent approximately every 3.5 seconds allowing for the collection of thousands of indents overnight. These indents are performed in a regular array, creating a map where every pixel corresponds to one indent. The collection strategy combines a piezo stage for high-resolution spatial accuracy across small areas, and a stepper motor geared stage to allow for large collection areas, albeit with relatively lower positional accuracy. As such, the overall array is formed of a collection of sub-arrays, herein named "bundles", within which the piezo stage controls position. The stepper motor movements are used to locate the centre of each bundle, allowing them to be stitched together. A graphical representation of this map-population strategy is shown in Figure 9.

For each indent the output data are values for maximum depth, load, modulus, hardness, and contact stiffness, instead of the conventional load-displacement curves collected in continuous stiffness measurements (CSM) [74], [75]. Though the load-displacement curve is not saved by the software in order to reduce file size, the modulus is calculated "on-the-fly" through the Oliver and Pharr method using a power-law fit on the unloading portion of the load-displacement curve [1]. The hardness is calculated through this stiffness-corrected contact depth using the same method [1]. This is in contrast to the more conventional CSM method used to validate measurements, possibly leading to the variations seen in these results.

Prior to discussions of analysis, the effect of thousands of indents on tip wear and calibration was considered. It is advisable for tip area functions to be re-calibrated regularly, either through direct means via AFM measurement, or indirect means via indentation in reference material [76]. In common practice, this would often occur after several hundred indents had been performed at most, a time-scale which could span several months. However, nanoindentation maps can often contain orders of magnitude larger number of indents, with no opportunity to calibrate during data collection. We experimentally verify if tip area functions obtained through calibration on reference materials persist throughout several tens of thousands of indents. To illustrate this, an indentation map of over 33,000 indents was performed on fused silica with a maximum load of 3 mN, spaced 1.5 μm apart. This corresponds to a depth/spacing ratio of ~1:10, with indents approximately ~150 nm deep. This was preceded and followed by two sets of nine CSM nanoindents to a target depth of 2000 nm. These were performed with an indentation strain rate of 0.05 $s^{-1}$, with a harmonic oscillation of 2 nm at 45 Hz, parameters used for all other CSM tests in this study. The first of these sets was used for tip area function calibration, using the software's default analytical method with parameters as set out in Table I.

Further to the tip calibration persistence test described above, a series of nanoindentation mapping tests were performed at depths of 100, 200 and 500 nm and compared to slower, conventional CSM tests at equivalent depths. The nanoindentation maps in these tests were conducted at a depth to spacing ratio of 1:10 (a discussion of this choice is given in the results section). In specimen 1, the nanoindentation map was performed at a fixed load of 3 mN, corresponding to an approximate depth of 200 nm, with an indent spacing of 2 μm. In specimen 2, the nanoindentation map was performed at a fixed load of 2.5 mN, corresponding to an approximate depth of 150 nm, with an indent spacing of 2 μm.

**Analysis Methodology**

The analysis method presented correlates spatially resolved property maps on a pixel-by-pixel basis. An experimental and analytical challenge lies in the correct registration of property maps obtained across a wide variety of means. In particular, aligning EBSD signal maps onto nanoindentation maps presents a challenge due to the marked difference in the method of signal collection. Though tilt correction is applied, EBSD collection is performed with electron beam incidence angle of 70 degrees, and a small angular deviation between the two reference frames on the specimen can lead to significant misalignment across maps [77]. In order to resolve this, geometrical transformations and interpolation are used to map all datasets onto the same grid of coordinates. The method presented takes the X and Y coordinates of the nanoindentation maps to be the most accurate coordinate frame.

The transformations are carried out individually, relating all other property maps to the nanoindentation map. Fiducial markers or easily identifiable features act as reference markers across the maps, and an affine transformation is performed on the property map to be registered. The use of the affine transformation assumes the presence of solely linear distortions in all other property maps, i.e. that all distortions are as a result of geometrical projections from collection angles. This implies that the distortion can be described as the sum of x and y translation, scaling, rotation, and shear alone. It is recognised that other, non-linear sources of distortion may be present. For example, mechanical drift during EBSD mapping could be non-linear. However, we observe that affine transformation accounts for the majority of discrepancy between mapping modalities used here at their current resolutions and collection timings. The best fit affine transformations required to map the positional EBSD and EPMA data onto the nanoindent reference axes were established by user identification of multiple corresponding points (at least four) within each dataset.

Following this, all property maps were interpolated in order to be scaled into an array the same size as the nanoindentation map. This often reduces the size of the dataset as nanoindentation mapping is not as high resolution as EBSD or EPMA, but retains the confidence that information obtained from correlations arises from real indents performed on the local material at that point. A "nearest-neighbour" interpolation was used for EBSD datasets to avoid interpolating grain boundaries as smooth transitions, while a "linear" interpolation was used for EPMA maps as we expect smooth diffusion gradients.

The same process can be applied to any further property map taken: the position information is transformed, and the property data interpolated appropriately such that a stack of maps is produced, each pixel containing a vector of property information for a given location. It is then possible to correlate individual property pairs by threading through the stack of maps.

A step by step guide is shown in the results of the first dataset. The code used for this analysis was written in MATLAB R2019a ® [78], with the code necessary to reproduce these results uploaded to a GitHub Repository named "XPCorrelate" under an MIT License: https://github.com/cmmagazz/XPCorrelate.

### Acknowledgements

CMM would like to acknowledge financial support from the EPSRC, Rolls-Royce plc, and the Royal Commission for the Exhibition of 1851. CMM would also like to extend gratitude to advisors at Rolls-Royce plc including Prof D Rugg, and Dr A Radecka, as well as the advice and services offered at the David Cockayne Centre for Electron Microscopy in the University of Oxford.

**Author contributions - CREDIT**

**CM Magazzeni:** Conceptualisation, Methodology, Software, Validation, Formal Analysis, Investigation, Data Curation, Writing, Visualisation. **HM Gardner:** Methodology, Investigation, Validation. **I Howe:** Conceptualisation, Software. **P Gopon:** Methodology, Investigation, Validation, Writing. **JC Waite:** Methodology, Validation. **D Rugg:** Supervision, Project Administration, Funding Acquisition. **DEJ Armstrong:** Conceptualisation, Project Administration, Funding Acquisition. **AJ Wilkinson:** Supervision, Conceptualisation, Formal Analysis, Writing, Project Administration, Funding Acquisition.

**Analysis Software**

**SUPPLEMENTARY INFORMATION**

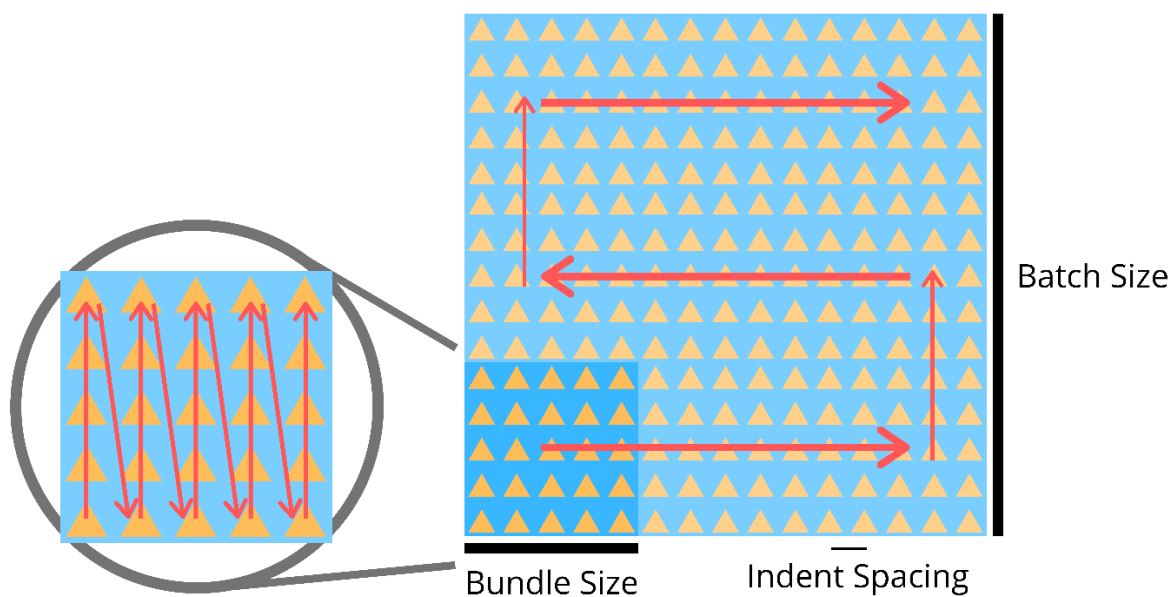

*Figure 9: Graphical representation of nanoindentation mapping strategy. Individual bundles, shown in the inset on the left, are populated line by line travelling along the y-axis. The map is populated with bundles in a serpentine motion, with the direction depending on the position of the starting point.*

# Comparison of conventional CSM nanoindentation vs high speed nanoindentation

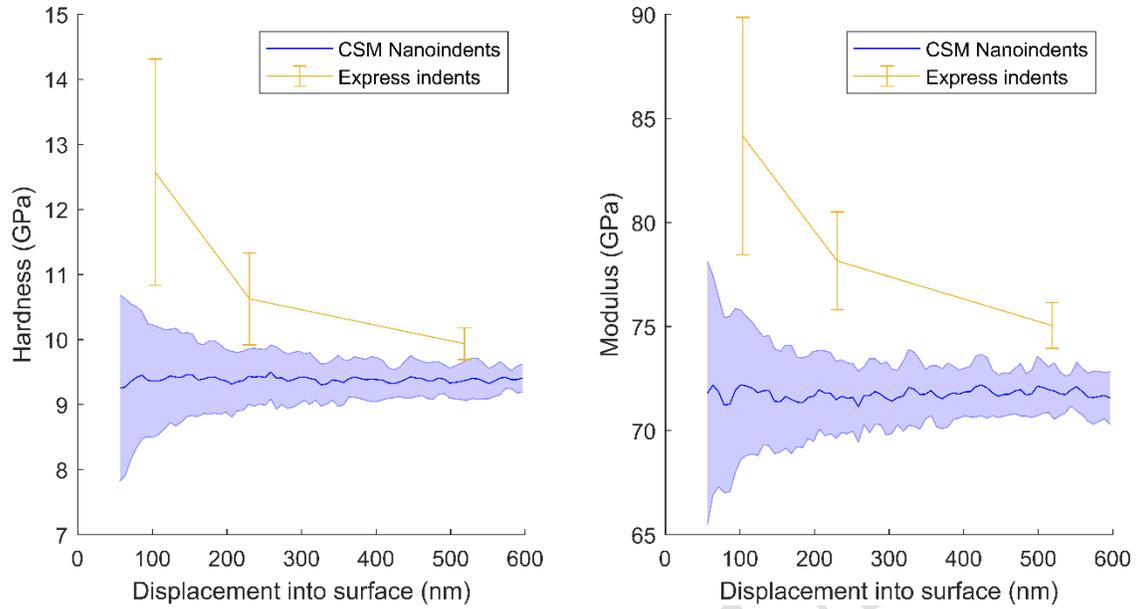

*Figure 10: Comparison of conventional CSM nanoindentation against high speed nanoindentation mapping, performed in fused silica. Mapping nanoindents were performed at a depth : spacing ratio of 1:10.*

# Comparison of conventional CSM nanoindentation vs high speed nanoindentation

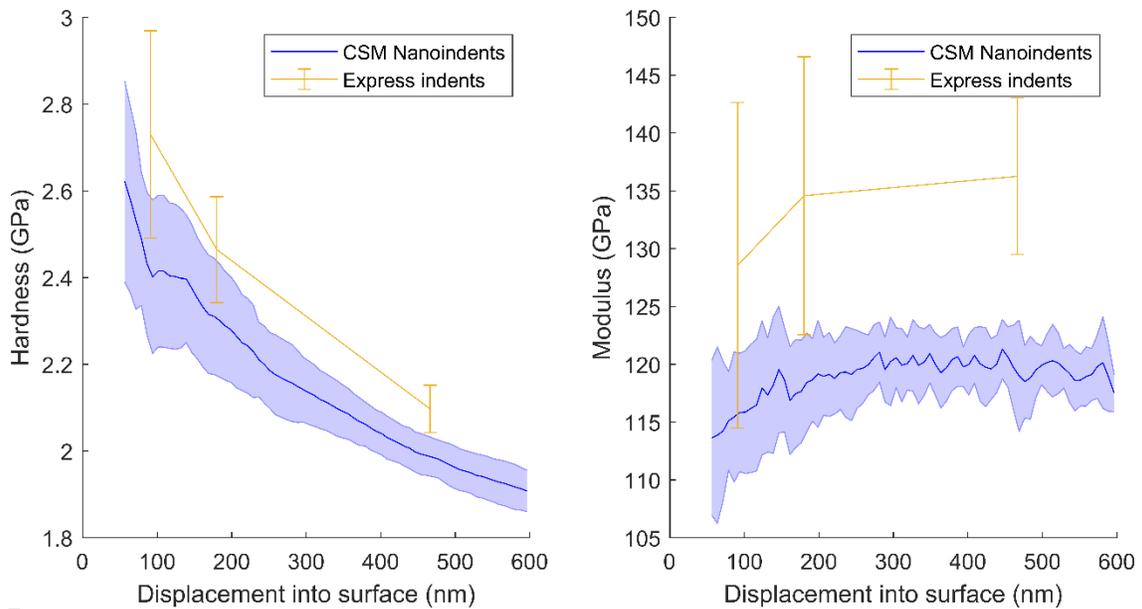

*Figure 11: Comparison of conventional CSM nanoindentation against high speed nanoindentation mapping, performed in Specimen 1 – CP titanium. Mapping nanoindents were performed at a depth : spacing ratio of 1:10.*

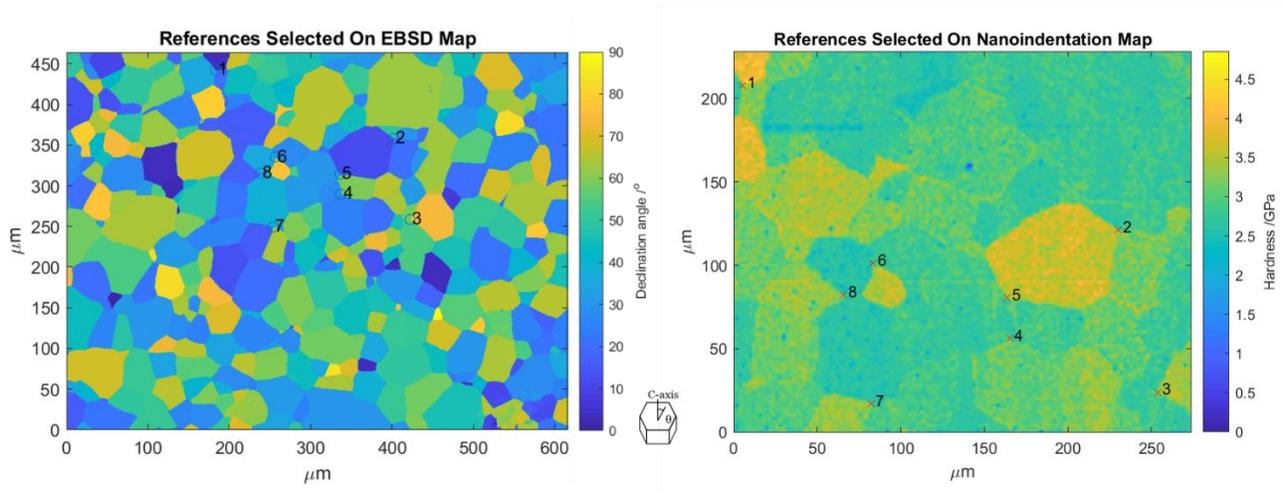

*Figure 12: EBSD (left) alongside the nanoindentation hardness map (right) highlighting the corresponding points in each map used for transforming the EBSD map.*

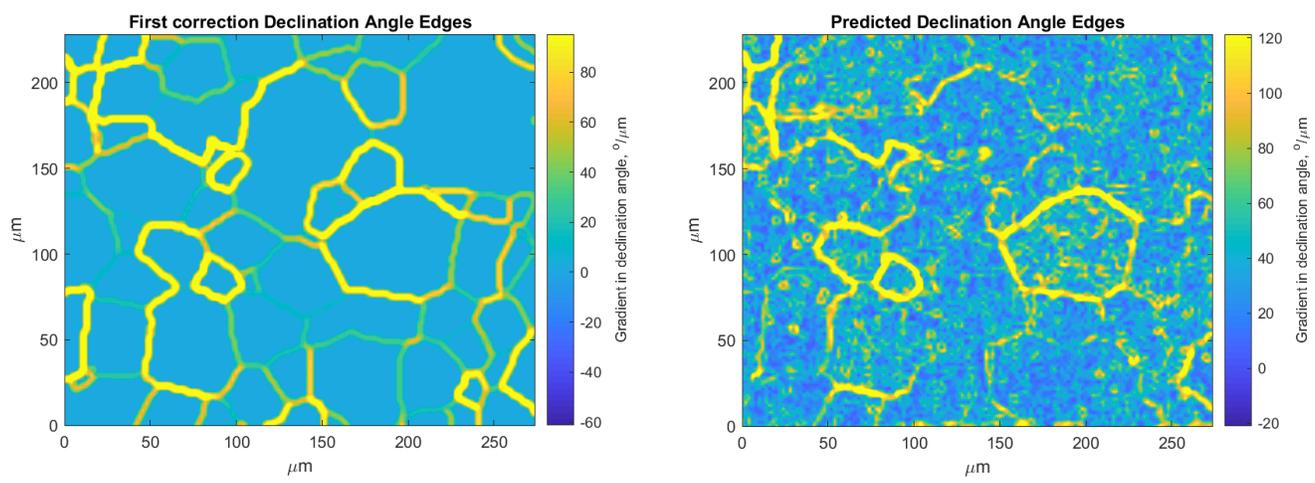

*Figure 13: EBSD declination angle edges, as aligned (left) compared with the simulated declination angle (right) calculated through the hardness relationship. These were obtained with a Sobel edge detection filter [57].*

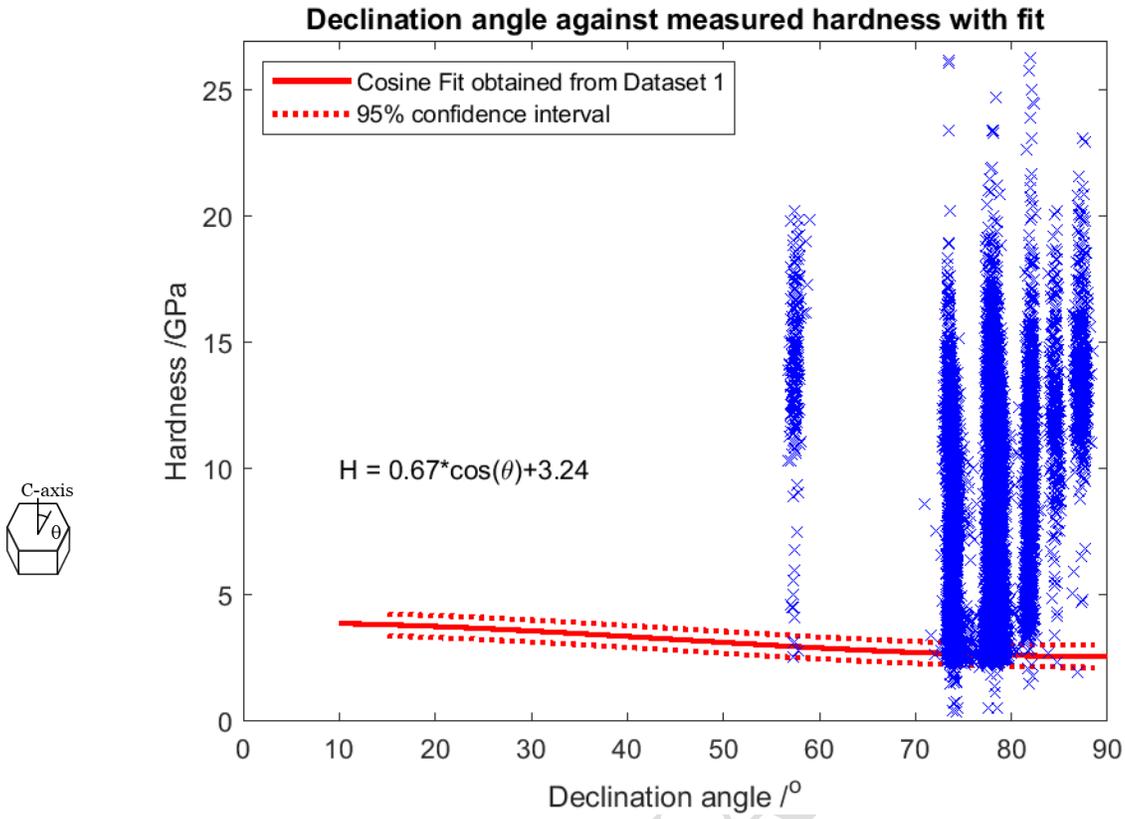

*Figure 14: Declination angle against measured nanoindentation hardness. A total of 26058 data points are shown.*

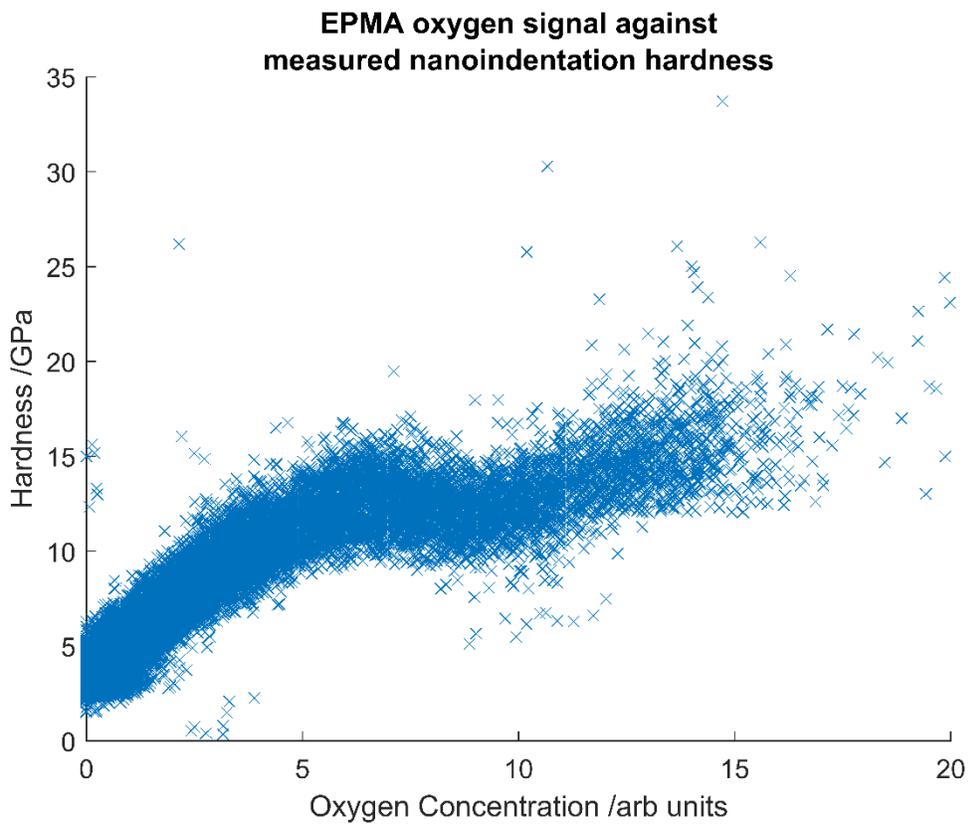

*Figure 15: Scatter plot of EPMA obtained oxygen signal against nanoindentation hardness. All points are shown.*

**Nanoindentation map highlighting points below 95% CI**

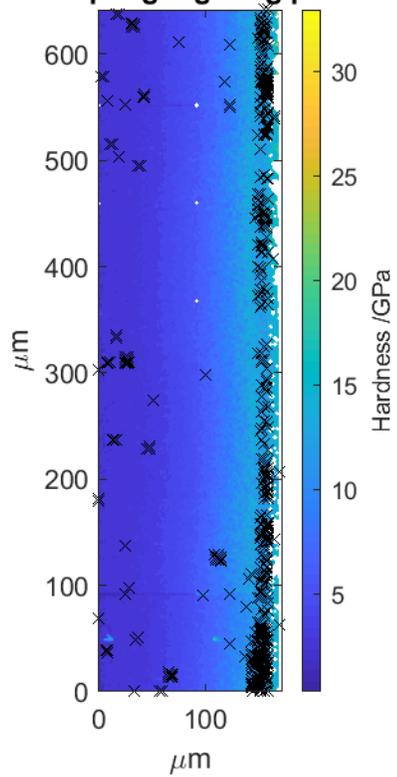

*Figure 16: Nanoindentation obtained hardness map with points lying below the 95% confidence interval highlighted with black crosses.*

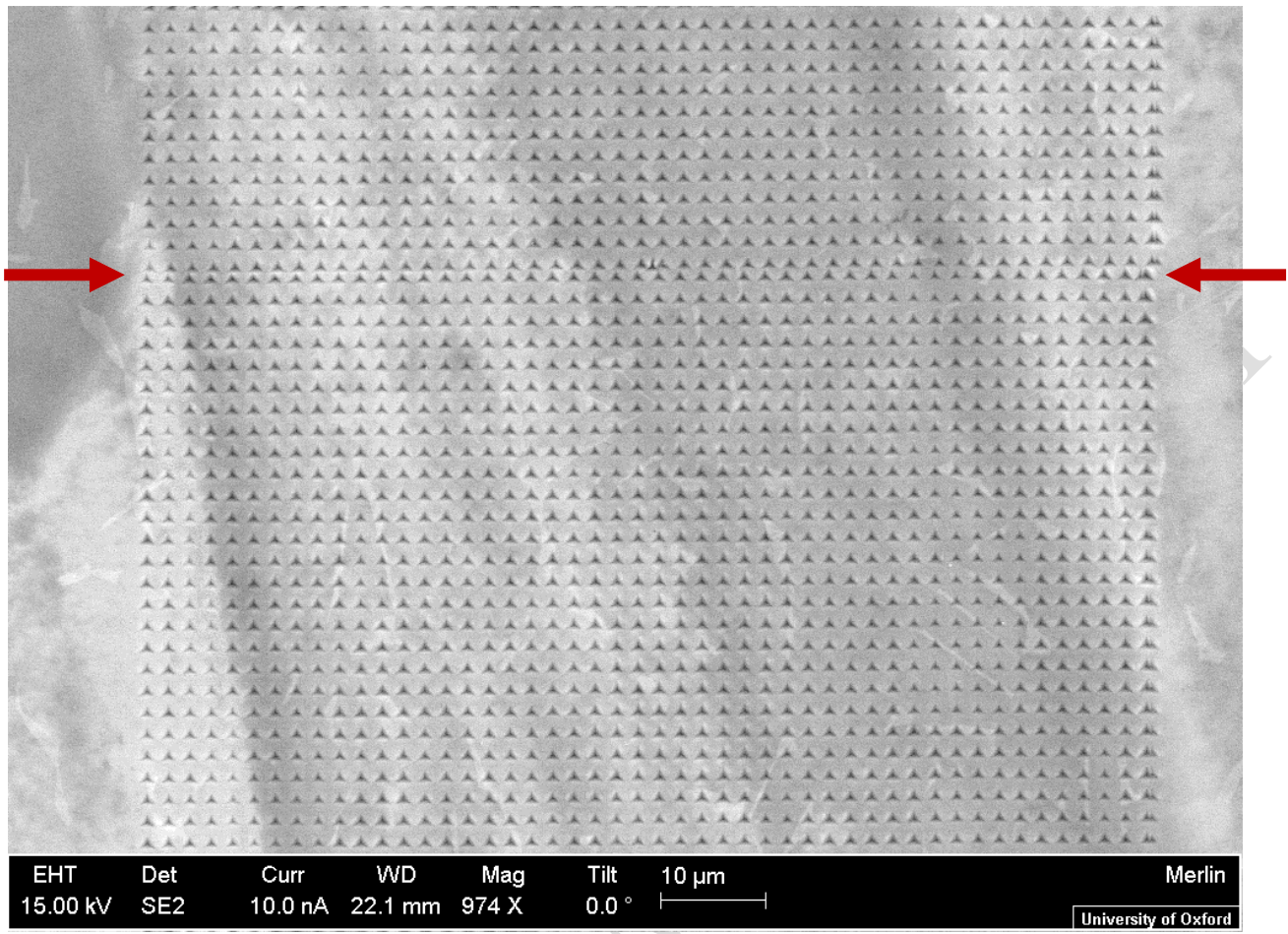

| EHT | Det | Curr | WD | Mag | Tilt | 10 μm | Merlin |
| 15.00 kV | SE2 | 10.0 nA | 22.1 mm | 974 X | 0.0 ° | | University of Oxford |

*Figure 17: SEM image of nanoindentation mapping. Image shows two bundles of 48x48 indents spaced 2 μm apart, with the border between them situated along the top of the image. A small offset between the bundles can be seen as a row of indents placed closer to each other than designed, highlighted with red arrows.*

**Tables**

*Table I: Parameters used during software calibration of tip area function through CSM indents on fused silica.*

| Parameter | Value / Range |
|---|---|
| Range of depth for area calculation | 50-2000nm |
| Nominal modulus of tested material | Fixed at 72 GPa |
| Range of depth for load frame stiffness calculation | 1500-2000nm |
| Coefficient count | 9, allowing negative |
| Lead term for area function | Unfixed |
| Analytical model: Beta and Epsilon | 1.00, 0.75 (default)[80] |